\documentclass[aps,prc,amsmath,amssymb,preprintnumbers,superscriptaddress,twocolumn,showpacs,floatfix]{revtex4}
\usepackage{epsfig,graphics,color}
\usepackage{graphicx}
\usepackage{dcolumn}
\usepackage{bm}
\usepackage{amsmath}

\newcommand \tie {{\it i.e.}}

\newcommand \ra  {\rightarrow}

\newcommand \vecr {\vec{r}}

\newcommand \e {\epsilon}

\newcommand \A {\alpha}

\newcommand \lc {\langle}
\newcommand \rc {\rangle}

\newcommand \D {\Delta}
\newcommand \sg {\sigma}

\newcommand \nt {\noindent}

\newcommand \bvec{\left( \begin{array}{c} }
\newcommand \evec{\end{array} \right)}

\newcommand \bea{\begin{eqnarray} }
\newcommand \eea{\end{eqnarray} }
\newcommand \nn {\nonumber}
\newcommand {\be} {\begin{equation}}
\newcommand {\ee} {\end{equation}}

\newcommand {\mbx} {\mbox{}}

\newcommand {\ata} {&\times&}

\begin{document}

\title{Jet modification in three dimensional fluid dynamics at next-to-leading twist}

\author{A. Majumder}
\affiliation{Department of Physics, Duke University, Durham, NC 27708, USA}
\author{C. Nonaka}
\affiliation{Department of Physics, Nagoya University, Nagoya 464-8602, Japan}
\author{S. A. Bass}
\affiliation{Department of Physics, Duke University, Durham, NC 27708, USA}


\date{\today}

\begin{abstract}
The modification of the single inclusive spectrum of high transverse momentum ($p_T $) pions emanating from an 
ultra-relativistic heavy-ion collision is investigated. 
The deconfined sector is modelled using a full three 
dimensional (3-D) ideal fluid dynamics simulation. 
Energy loss of high $p_T$ partons and the ensuing modification of  their fragmentation 
is calculated within perturbative QCD 
at next-to-leading twist, where the magnitude of the higher twist contribution is modulated by the entropy density 
extracted from the 3-D fluid dynamics simulation. The nuclear modification factor ($R_{AA}$) for  pions 
with a $p_T  \geq 8$ GeV as a function of 
centrality as well as with respect to the reaction plane is calculated. The magnitude of contributions to the differential 
$R_{AA}$  within small angular ranges, 
from various depths in the dense matter is extracted from the calculation
and demonstrate the correlation of the length integrated density and the $R_{AA}$ from a given depth. 
The significance of  the 
mixed and hadronic phase to the overall magnitude of  energy loss are explored.    
\end{abstract}

\maketitle

Current experiments at the Relativistic Heavy Ion Collider (RHIC) have 
established two fundamental properties of the dense partonic matter 
produced in the collision of heavy ions~\cite{RHIC_Whitepapers}. 
The produced matter achieves 
local thermal equilibrium rapidly and is endowed with a very low 
viscosity; as a result, the dense partonic matter demonstrates almost 
ideal hydro-dynamic expansion~\cite{Teaney:2003kp}. 
Secondly, the propagation of high 
transverse momentum partons, produced in 
the early hard scatterings, is systematically impeded.  The 
spectrum of  high transverse momentum hadrons with $p_T \geq 8$~GeV 
shows a reduction of almost a factor of 5 in central $Au$-$Au$ collisions compared to that 
from the spectrum of  such particles in elementary nucleon nucleon encounters 
enhanced by the number of expected binary collisions~\cite{Adler:2003qi}.

In most estimates from \emph{jet-quenching} calculations~\cite{jet_quenching,Salgado:2003gb}, 
there exists no real dynamical model of the medium~\cite{jet_geometry}; in the best of circumstances, a time dependent, 
three dimensional, functional form $\rho(x,y,z,\tau) \sim \rho_0(x,y,z)(\tau_0/\tau)$, with the initial 
density estimated from a wounded-nucleon model, is used. 
There have been extentions involving a 1-D hydrodynamical model~\cite{Renk:2006qg}, however 
such calculations cannot address the centrality dependence of the modification.  
The first attempt involving a realistic model of the medium~\cite{Hirano:2002sc}
estimated the effects of 3-D expansion on the uncharged pion spectrum and the  $R_{AA}$.
However, this approach treated the
energy loss of jets in a rather simplified  manner.
Detailed 3-D simulations of the medium have recently become available in Ref.~\cite{Nonaka:2006yn}, 
where the various parameters of the simulation were fitted by comparison with 
experimental data in the soft sector. A study of the modification of jets in this medium, using the quenching weights 
of Ref.~\cite{Salgado:2003gb}, was 
carried out in Ref.~\cite{Renk:2006sx}. However, such calculations, along with others have tended 
to reformulate the effect of a 3-D expanding medium in terms of an effective length. 
In the current effort, the medium modified fragmentation functions of  high  $p_T$ pions 
will be computed in the higher-twist energy loss formalism~\cite{guowang}. The 
3-D expanding plasma of Ref.~\cite{Nonaka:2006yn}  will furnish the space-time 
dependence of the two-gluon matrix element (related to the well known 
transport coefficient $\hat{q}$), which parameterizes the sensitivity of the 
jet-modification equations to the properties of the medium. The calculation of the contributions which lead to 
the Landau Pomeranchuck Migdal (LPM) interference will be carried out directly in such a medium without 
recourse to an effective length.     
In the following, the salient features of the 3-D fluid dynamical 
simulations and the ensuing calculation of  the nuclear modification factor at 
next-to-leading twist will be briefly reviewed. 

Ideal Relativistic Fluid Dynamics (RFD, see e.g.~\cite{Bjorken:1982qr}) 
is the study of matter where the precise degrees of freedom are unimportant to the evolution 
of the extensive variables of the system. It is ideally suited for the high-density strongly coupled 
phase of heavy-ion reactions at RHIC,
but breaks down in the later, dilute, stages of the
reaction when the mean free paths of the hadrons become
large and flavor degrees of freedom
are important. 
Relativistic Fluid Dynamics is based on the solution of the 
four equations of energy momentum conservation, 
 \begin{equation}
\partial_\mu T^{\mu \nu} (x) = 0,
\label{Eq-rhydro}
\end{equation}
where $T^{\mu \nu}$ is the energy momentum tensor, given by
\begin{equation}
T^{\mu \nu}=(\epsilon + p) U^{\mu} U^{\nu} - p g^{\mu \nu}.
\end{equation}
Here, $\epsilon$, $p$, $U$ and $g^{\mu \nu}$ are the local energy density,
pressure, four velocity and metric tensor, respectively at $x$. 
The relativistic hydrodynamic equation [Eq.~(\ref{Eq-rhydro})]
is solved numerically using baryon number $n_B$ conservation
\begin{equation}
\partial_\mu (n_B (T,\mu) U^\mu)=0,
\end{equation}
as a constraint and closing the resulting set of partial
differential equations by specifying an equation of state (EoS):
$\epsilon = \epsilon(p)$.
In the ideal fluid approximation
(i.e. neglecting off-equilibrium effects), once the initial conditions
for the calculation have been fixed,
the EoS is the {\em only}
input to the equations of motion and relates directly to the 
properties
of the matter under consideration. 
As a result, even striking phenomena occurring  within systems in local equilibrium, such as 
phase transitions may be easily incorporated into the bulk dynamics of the full system.
Ideally, either the initial conditions or the
EoS should be determined beforehand by an ab-initio calculation (e.g. the EoS may be 
obtained from a lattice-QCD calculation), in which case, a fit to the data would allow
for the determination of the other quantity.

We assume that hydrodynamic expansion starts at
$\tau_0=0.6$ fm. Initial energy density and
baryon number density are parameterized by
\begin{eqnarray}
\epsilon(x,y,\eta)& =& \epsilon_{\rm max}W(x,y;b)H(\eta),
\nonumber \\
n_B(x,y,\eta)& = & n_{B{\rm max}}W(x,y;b)H(\eta),
\end{eqnarray}
where $b$ and  $\epsilon_{\rm max}$ ($n_{B{\rm max}}$) are
the impact parameter and the maximum value of energy density
(baryon number density), respectively.
$W(x,y;b)$ is given by a combination of wounded nuclear model and
binary collision model \cite{Kolb:2001qz} and  $H(\eta)$ is given
by $\displaystyle
H(\eta)=\exp \left [ - (|\eta|-\eta_0)^2/2 \sigma_\eta^2 \cdot
\theta ( |\eta| - \eta_0 ) \right ]$.
RFD has been very successful in describing soft matter
properties at RHIC, especially
collective flow effects and particle spectra~\cite{Nonaka:2006yn,Kolb:2003dz}.
All parameters of our hydrodynamic evolution ~\cite{Nonaka:2006yn}
have been fixed
by a fit to the soft sector (elliptic flow, pseudo-rapidity
distributions and low-$p_T$ single particle spectra), therefore
providing us with a fully determined 
medium evolution for the hard probes to propagate through.
While different descriptions for the hadronic phase were presented in Ref.~\cite{Nonaka:2006yn}, 
in this calculation, the entire evolution of both the deconfined and hadronic phase will be assumed to 
be describable by the RFD simulation.


Assuming a factorization of initial and final state effects, the differential cross-section 
for the production of a high $p_T$ hadron at midrapidity from the impact of 
two nuclei $A$ and $B$ is 
obtained as a convolution of initial nuclear structure functions, 
$G^A_a(x_a,Q^2),  G^B_b(x_b,Q^2)$ (shadowing functions are taken from Ref.~\cite{Li:2001xa}),
a hard cross section to produce a high $p_T$ parton ($d \hat{\sg}/ d \hat{t}$) and a medium modified fragmentation function. 
The entire effect of  final state jet modification is encoded in the medium modified 
fragmentation function $\tilde{D}(z,Q^2)$ expressed as the sum of the leading 
twist vacuum fragmentation 
function~\cite{bin95} and a correction from re-scattering of the struck 
quark  in the medium \tie, $ \tilde{D} = D + \D D$~\cite{guowang}. 
In the collinear limit, the modification is computed by isolating corrections, suppressed by powers of $Q^2$, 
which are enhanced by the length of the medium~\cite{lqs}. At next-to-leading twist, the correction for the 
fragmentation of a quark has the expression (generalized from deep-inelastic scattering (DIS)~\cite{guowang,maj04e}), 

\begin{eqnarray}
\!\!\!\D D(z,Q^2 ,\vecr) &=& \frac{\A_s}{2\pi} \int \frac{dl_\perp^2}{l_\perp^2} 
\int \frac{dy}{y} P_{q \ra i}(y)  2\pi \A_s C_A           \nn \\
\ata T^{M}(\vec{b},\vecr,x_a,x_b,y,l_\perp)D_i \left( z/y , Q^2 \right) \nn \\ 
\ata \left[\frac{\mbox{\Large}}{\mbox{\Large}} \right.  
\mbx\!\!\!\int \!\!d^2 b d^2 r  
t_A(\vec{r}+\vec{b}/2) t_B(\vec{r} - \vec{b}/2) \nn \\ 
 \ata \left.  l_\perp^2 N_c G^A_a(x_a) G^B_b(x_b) 
\frac{d \hat{\sg}}{d \hat{t}} \right] ^{-1}  +  v.c.  \label{delta_D}
\end{eqnarray}

\nt
In the above equation, $l_\perp$ is the transverse momentum of the radiated gluon (quark) which 
leaves a momentum fraction $y$ in the quark (gluon) denoted as parton $i$ [$P_{q\ra i}(y)$ is the splitting function for 
this process] which then fragments leading to 
the detected hadron. The factors $t_A(\vec{r}),t_B(\vec{r})$ 
are the thickness functions of nuclei $A$ and $B$ at the transverse location $\vec{r}\equiv(x,y)$.
The $v.c.$ refers to  
virtual corrections. 
The factor $T^{M}$, which has its origin in higher twist matrix elements of the quark, may 
be expressed as, 

\begin{eqnarray}
T^{M} \!\!\!&=&   \!\!\int\!\! d^2b d^2 r t_A(\vecr + \vec{b}/2)  t_B(\vecr -\vec{b}/2)
G^A_a(x_a) G^B_b(x_b)  \frac{d \hat{\sg}}{d \hat{t}}   \nn \\
 \ata \int_0^{\zeta_{max}} \!\!\!\!\!\! d \zeta x_g \rho_g(x_g,\hat{n}\zeta + \vecr) (2 - 2 \cos(\eta_L \zeta))  . \label{TM}
\end{eqnarray}

\nt
In the above equation, the factor 
$\eta_L = l_\perp^2/ [2 \hat{p}_T y(1-y)]$. 
Where, $\hat{p}_T$ represents the transverse 
momentum of the produced parent jet. 

The jet direction ($\hat{n}$) is chosen as the 
$x$-axis in the transverse $(x,y)$ plane ($z$-direction is set by the beam line); 
the angle of the reaction plane 
vector $\vec{b}$ is measured with respect to this direction.  The distance travelled 
by the jet in this direction prior to 
scattering off a gluon is denoted as $\zeta$.  The gluon's forward (towards the jet) momentum 
fraction in denoted as $x_g$. 
In what follows, the nuclear modification of the single inclusive spectrum, both integrated and as a function of the 
angle with respect to the reaction plane will be computed. The angle integrated modification, the $R_{AA}$, is defined as,

\begin{eqnarray}
R_{AA} &=& \frac{\frac{d \sigma^{AA}}{dy d^2 p_T} }
{T_{AA} (b_{min},b_{max})\frac{d \sigma^{pp}(p_T,y)}{dy d^2 p_T}}, \label{raa}
\end{eqnarray} 
\nt
where, $T_{AA}(b_{min},b_{max})$ is related to the mean number of binary nucleon nucleon 
encounters in the range of impact parameter chosen ($\lc N_{bin} \rc$), the geometric $A$-$A$ cross section [$\sg^{AA}_{Geo}(b_{min},b_{max})$]  and the 
inelastic $p$-$p$ cross section ($\sg_{pp}^{in}$) as 
$T_{AA}(b_{min},b_{max}) \simeq \lc N_{bin} \rc \sg^{AA}_{Geo}(b_{min},b_{max})/\sg_{pp}^{in}$. 

The phenomenological input that is required to understand the variation of the nuclear modification 
factor with centrality or $p_T$ is the space-time dependence of the gluon density $\rho(x,y,z,\tau) = x_g \rho_g(x_g,x,y,z,\tau)$. 
This density is directly proportional to the 
transport coefficient 
$\hat{q}_R$ (for a jet parton in representation $R$)~\cite{Majumder:2006we,Baier:1996sk}.
%

In the remainder, the $z$-coordinate will be set to zero as the interest will remain on midrapidity 
observables. The gluon density at a given location depends on the phase of matter at that location and 
is in general assumed to depend on the number of degrees of freedom prevalent in the excited matter 
at that location. The exact degrees of freedom in the deconfined phase are as yet unknown~\cite{Shuryak:2004tx}. 
The partonic density as seen by the jet within each degree of freedom is also unclear. For this first attempt at higher 
twist energy loss in a hydrodynamically expanding medium, we invoke the very simple ansatz,  

\bea
\hat{q}(x,y,z,\tau) &=& \hat{q}_{0}  
\frac{  \gamma_\perp (x,y,z,\tau) T^3(x,y,z,\tau) }{T_0^3} \label{qhat_pres} \\
\mbx \times && \!\!\!\!\!\!\!\!\!\!\!\!\left[ R(x,y,z,\tau) + c_{HG} \left\{  1 - R(x,y,z,\tau) \right\}  \right] , \nn
\eea
\nt
where, $T(x,y,z,\tau)$, $\gamma_\perp (x,y,z,\tau)$ and $R(x,y,z,\tau)$ represent the temperature, 
flow transverse to the jet and the volume fraction in the plasma phase at the space-time 
point $x,y,z,\tau$. It is this information that 
is extracted from the RFD simulation. The factors $\hat{q}_0,T_0$ represent the maximum 
$\hat{q}$ and temperature achieved in the simulation; in this particular version of RFD, $T_0 = 0.405$ GeV and 
$\hat{q}_0$ is a fit parameter adjusted to fit one data point of the $R_{AA}$, at one centrality. 
The factor $c_{HG}$ accounts for the 
fact that the partonic density per degree of freedom in the hadron gas phase may be 
different from that in the QGP phase and plots with different choices of $c_{HG}$ will be 
presented.  It should be pointed out that contrary to the (pre-)hadronic scattering picture of 
Ref.~\cite{Gallmeister:2002us},  the energy loss of jets in the hadronic 
phase is considered here as a completely partonic process~\cite{guowang}, where 
the jets fragment into hadrons outside the medium. 

\begin{figure}[htbp]
\resizebox{2.2in}{2.0in}{\includegraphics[0.75in,0.5in][4.75in,4.5in]{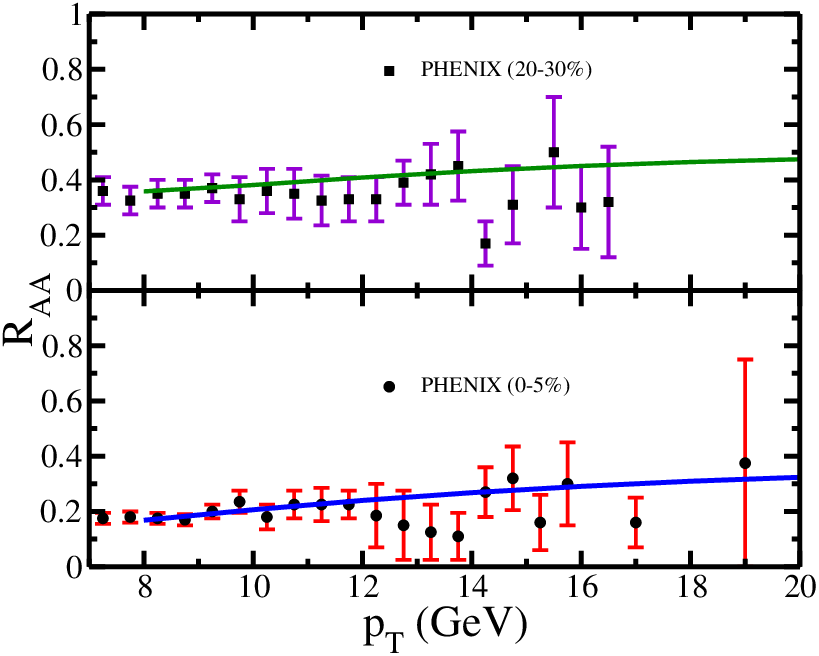}}
\caption{(Color online) $R_{AA}$ (with $\hat{q}_0 \simeq 1.3$ GeV$^2/$fm and $c_{HG}=1$) in $Au$-$Au$ collisions at 0-5\% (blue line) 
and 20-30\% (green line) centrality compared to data from PHENIX~\cite{Shimomura:2005en}.  }
\label{fig1}
\end{figure}

The above prescription of  assuming $\hat{q}$ to be a linear function of $T^3$ 
may be referred to as the thermal prescription. 
In the energy density prescription used in Ref.~\cite{Renk:2006sx}, one assumes $\hat{q}$ 
to be a linear function of $\e^{3/4}$. Alternatively, one may also assume that $\hat{q}$ is 
a linear function of the entropy density $s$. Relationships may be drawn between these 
various schemes with a knowledge of the EoS. For instance, in a plasma phase which is populated 
by massless partons, the entropy density \emph{is} a linear function of $T^3$ and the 
two prescriptions are identical.  
This breaks down in the hadronic phase as the entropy density is no longer a simple linear function 
of $T^3$.  
Due to the variety of unknowns regarding the $\hat{q}$ in an excited hadronic resonance  and as a result, its parametric 
dependence in the hadron resonance phase, we persist with the simple prescription of  Eq.~\eqref{qhat_pres}.  

As mentioned above, the overall fit parameter $\hat{q}_0$ is tuned to fit one experimental data point, 
at one centrality and $p_T$. For the current effort, the fit parameter is set by requiring that the 
$R_{AA}$ at $p_T=10$ GeV in the most central event ($0$-$5\%$ centrality) is $0.2$. For a 
$c_{HG}=1$ this corresponds to a $\hat{q}_0 \simeq 1.3$ GeV$^2/$fm for gluon jets. 
With the value of $\hat{q}_0$  and $c_{HG}$ fixed, the variation of $R_{AA}$ as a function of 
$p_T$ and centrality of the collision are predictions. These are presented in 
Fig.~\ref{fig1} for two centralities (0-5\% and 20-30\%) in comparison with data from 
PHENIX~\cite{Shimomura:2005en}. An important caveat to the estimates of $\hat{q}$ are the
estimates of the plasma temperature from the very early phase of the evolution. 
Hard jets are produced very early ($t<0.1$ fm/c)
in the evolution of the system. However, fits to the spectrum of soft particles indicate 
that a locally equilibrated matter does not arise before 0.6 fm/c. This is the starting 
point of the RFD simulations used. In the regime between 0.1 to 0.6 fm/c, estimates of $\hat{q}$ 
are open to speculation. In this first effort, the initial $\hat{q}$ is assumed to be held 
constant from the time the jets are produced to 0.6 fm/c. In future efforts, variations 
of this form will be attempted.

\begin{figure}[htbp]
\resizebox{2.5in}{2.0in}{\includegraphics[0.5in,0.5in][4.5in,4.5in]{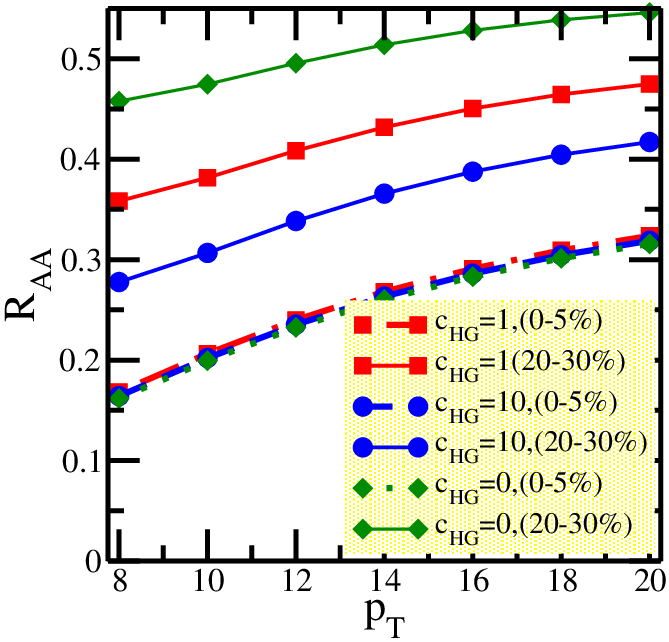}}
\caption{(Color online) $R_{AA}$ as a function of centrality for different choices of  
$c_{HG} = \hat{q}_{HG}/T^3$ in the hadronic phase. See text for details.}
\label{fig2}
\end{figure}

In the RFD 
simulations used, an equation of state with a first order phase transition is included. 
As a result, such simulations include a prominent hadronic and mixed phase. 
In Fig.~\ref{fig1}., the factor $c_{HG}=1$ meaning that the $\hat{q}$ per  unit of 
$T^3$ in the deconfined phase is similar to that in the hadron gas phase. 
From the equation of state used, one may discern the approximate and simple relation: $s_{QGP} \sim 5 s_{HG}$,  between the 
entropy densities in the deconfined phase  ($s_{QGP}$)  and the hadronic phase ($s_{HG}$), in the vicinity of the 
phase transition. 
This implies that a $c_{HG} = 1$, places the ratio $\hat{q}/s$, \tie, 
the quenching strength per unit of entropy, in the hadronic phase to be five times larger than that 
in the QGP phase. Due to the error bars in the data, a range of values of $c_{HG}$ is 
allowed; however, certain  extremes may be ruled out.

\begin{figure}[htbp]
\resizebox{2.5in}{2.0in}{\includegraphics[0.5in,0.5in][4.5in,4.7in]{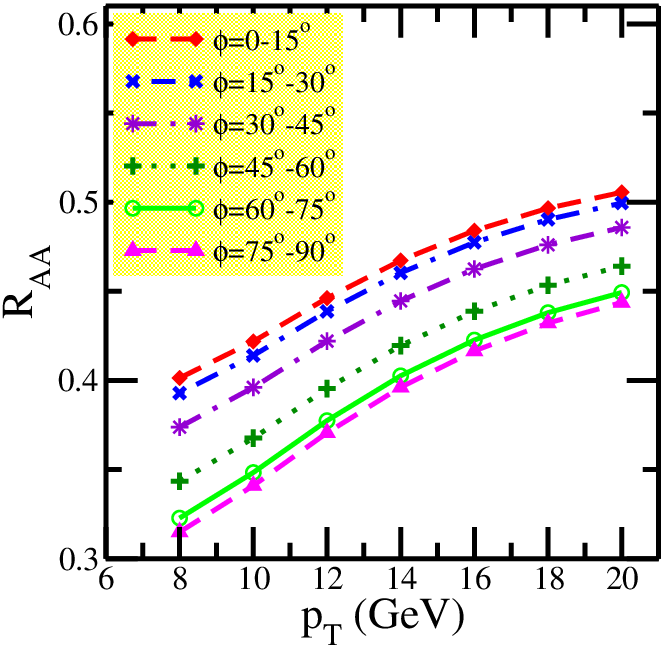}}
\caption{(Color online) $R_{AA}$ as a function of the reaction plane for $Au$-$Au$ collisions 
at $20-30\%$ centrality.}
\label{fig3}
\end{figure}

As an illustration, the $R_{AA}$ 
for the same centralities as in Fig.~\ref{fig1} are plotted for two other cases of $c_{HG} = 0$ and 
$c_{HG} = 10$ in Fig.~\ref{fig2}. In each case, the overall normalization factor $\hat{q}_0$ is
re-adjusted to the standard condition that the $R_{AA}$ for a pion $p_T= 10$ GeV and at a centrality of 0-5\% 
be 0.2. This is the reason that all three plots for the three different choices of $c_{HG}$ at 0-5\% centrality 
essentially lie on top of one another. However, the $R_{AA}$ for the  semi-central case (20-30\%) shows  noticeable 
variations. The red squares correspond to the case of Fig.~\ref{fig1}. The case $c_{HG} =0$, represented 
by the green diamonds, represents the case of no quenching in the hadronic phase. Due to the proportionately 
larger presence of the hadronic phase in more peripheral collisions, this case, understandably demonstrates 
less modification~\footnote{In the RFD simulations, no attempt is made to incorporate the fact 
that central collisions endure a longer lifetime and hence freeze-out later than peripheral collisions; simulations for all centralities 
are run up to 10 fm/c and then frozen out.}. 
 The other extreme 
is with a $c_{HG} = 10$, where 
the $\hat{q}$ in units of $T^3$ is an order of magnitude larger in the hadronic phase compared 
to the deconfined phase. As would be expected, the results show an opposite trend: 
the centrality dependence of the $R_{AA}$ is under-predicted.

In semi-central events, the average path length travelled by jets in the reaction plane are 
different from the path lengths travelled out of plane. This should lead to the occurrence of an 
azimuthally dependent nuclear modification factor. 
Full three dimensional RFD simulations capable of 
reproducing the observed azimuthal asymmetry in the low $p_T$ spectrum require 
that the initial gluon density profile display a considerable azimuthal eccentricity in 
transverse space. 
The hard scatterings, which lead to the production of back-to-back jets tend to produce them 
isotropically. As a result, a sizeable azimuthal asymmetry in the nuclear modification factor 
demonstrates the sensitivity of this observable to the early density profile of the QGP.
As an illustration of this effect, the nuclear 
modification factor at an angle $0^\circ \leq \phi \leq 90^\circ$ with respect to the 
reaction plane, in angular bins of $15^\circ $\cite{Majumder:2006we}, 
is plotted in Fig.~\ref{fig3}.  One notes, that 
the $R_{AA}$ in plane shows the least modification, which grows as the 
angular bin is turned away from the reaction plane. As a result, it may be argued that 
the initial profile of the produced matter plays an important role in the determination of the 
final nuclear modification factor.  The rate of change of the $R_{AA}$  as a function of  $\phi$ 
shows the interesting pattern that is slow at the terminal 
points and maximal in the middle of the angular range (\tie, $\phi \sim 45^\circ$).
The results obtained from this analysis (and especially the symmetric behavior of the 
modification as a function of angle with respect to the reaction plane) show an identical pattern as that of the experimental 
measurements of the $R_{AA}$ versus the reaction plane at lower $p_T$, as reported in Ref.~\cite{Adler:2006bw}.
The shallow rise in the $R_{AA}$ as a function of the $p_T$ of the detected hadron 
is demonstrated by the nuclear modification factor in each angular bin: this property is thus 
independent of the density of the matter probed by the jet. 

\begin{figure}[htbp]
\resizebox{2.5in}{2.0in}{\includegraphics[0.5in,0.5in][4.5in,4.7in]{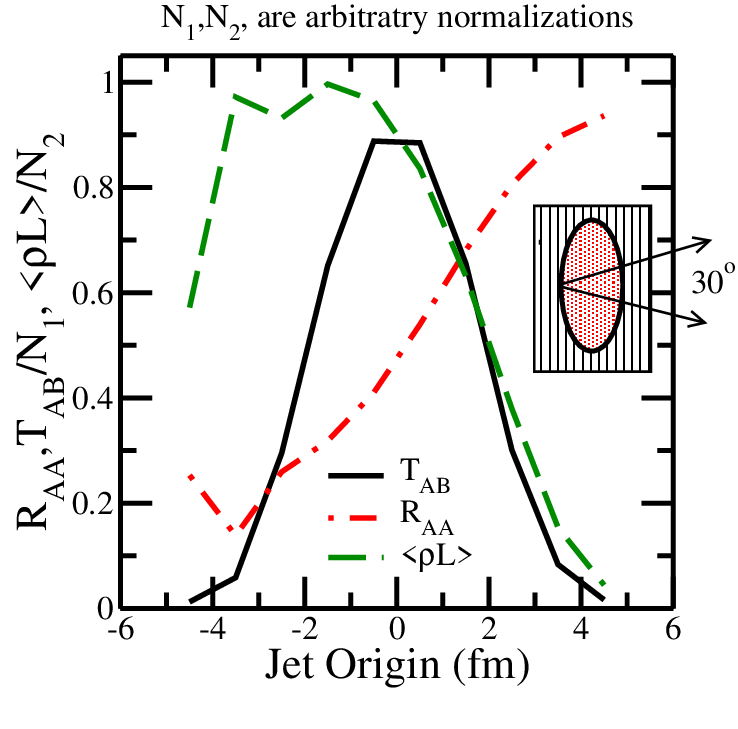}}
\caption{(Color online) $R_{AA}$ as a function of the location along the impact parameter from 
longitudinal strips perpendicular to $\vec{b}$. Also plotted in the number of initial binary 
nucleon-nucleon collisions and the length integral of the time dependent density weighted with the 
number of binary collisions $\lc \int dl \rho(x,y,z,t) \rc$ arbitrarily normalized to fit on the 
same plot. }
\label{fig4}
\end{figure}

The existence of a calculation scheme for the modification of jets in a  three dimensional 
dynamically evolving medium allows for a more differential analysis of jet quenching in 
medium. An an example, we compute the $R_{AA}$ as a function of location in the reaction 
plane of the 20-30\% centrality collisions. 
The focus is on the $R_{AA}$ in the angular bin of $0$-$15^\circ$ with respect to the 
reaction plane, this essentially corresponds to a $30^\circ$ opening angle around the 
impact parameter $\vec{b}$. The tansverse plane in then divided into ten longitudinal 
strips of width 1~fm perpendicular to $\vec{b}$ 
(as illustrated in the inset in Fig.~\ref{fig4}). The $R_{AA}$ for 
jets originating in a given strip is then plotted (red dot-dashed line) as a function of the location of the mid-point 
of the strip in the reaction plane.  Also plotted is the initial number of binary nucleon-nucleon 
scatterings in each of the strips (black solid line).
Plotted along with this is the averaged 
length integrated density $\rho$ [which is directly proportional to $\hat{q}$, see Ref.~\cite{Majumder:2006we,Baier:1996sk} ]
as experienced by a jet originating in each strip (green dashed line). This is 
defined as the following quantity 
\bea
\lc \rho L \rc &=& \int d^2 b d^2 r t_{A} (\vecr + \vec{b}/2) t_{B} (\vecr - \vec{b}/2 ) \\  
\ata \int d \phi_{\hat{n}} d\zeta \rho(\vecr + \zeta \hat{n},\tau=\zeta) \nn \\ 
\ata \left[ \int d^2 b d^2 r t_{A} (\vecr + \vec{b}/2) t_{B} (\vecr - \vec{b}/2)  \right]^{-1}. \nn
\eea 
\nt
In the above equation, the  density $\rho$ is evaluated at the location $\vecr + \hat{n} \zeta$ and 
at time $\zeta$, where $\vecr$ is the location of the jet vertex, $\zeta$ is the time elapsed and $\hat{n}$ 
is the direction of propagation of the jet.  The range of $\phi_{\hat{n}}$ is from $-15^\circ$ to $15^\circ$, while $\vecr$
ranges within the specified strip. Thus, 
$\lc \rho L \rc$ represents the length integrated density as experienced by a jet on average in the given 
strip.  
As may be noted from Fig.~\ref{fig4}, the $R_{AA}$ in a given strip is 
anticorrelated with the mean integrated density experienced by the jet on its way out of the 
medium. The $R_{AA}$ tends to show a small rise in the strip farthest from exit (note that this is 
correlated with a drop in $\lc \rho L \rc$), this is due to the fact that a considerable fraction of 
jets that originate on this strip assume tangential paths of exit that do not pass through the center of the
dense system produced.  As one moves towards the middle of the impact parameter, a large fraction 
of jets, due to the angular cut, are forced to pass through the densest part of the matter and as a result, the
$R_{AA}$ first drops and then rises.

In this letter, higher twist calculations of energy loss have been extended to include realistic 
medium evolution as stipulated by a 3-D hydrodynamical simulation. The nuclear modification 
factor at different centralities as a function of the $p_T$ of the detected hadron has been 
calculated and demonstrates very good agreement with experimental data. The variation of the 
$R_{AA}$  with centrality shows a noticeable sensitivity to the presence of a hadronic phase. 
This may be used to place bounds on the ratio between the $\hat{q}$ in the hadronic phase compared to the 
partonic phase. 
More differential observables, such as the 
$R_{AA}$  versus the reaction plane are calculated and show qualitative agreement with the data 
at a somewhat lower $p_T$. These results demonstrate the sensitivity of the $R_{AA}$ at a given 
angle with respect to the reaction plane to the initial profile of the 
produced matter. The nuclear 
modification factor as a function of the depth in the medium is shown to be anticorrelated
with the length integrated density as experienced by a jet on average. The magnitude of 
contributions to the $R_{AA}$  from different depths demonstrate the diminished surface 
bias in semi-central collisions.

{\it Acknowledgments:} The authors thank B. M\"{u}ller for insightful discussions. 
This work is  supported in part by a grant 
from the  U.~S.~D.~O.~E. (DE-FG02-03ER41239-0).

\end{document}